\documentclass[conference]{IEEEtran}
\IEEEoverridecommandlockouts

\usepackage{cite}
\usepackage{amsmath,amssymb,amsfonts}
\usepackage{algorithmic}
\usepackage{graphicx}
\usepackage{textcomp}
\usepackage{verbatim}
\usepackage{bm}  
\usepackage{subfigure}
\usepackage{xcolor}

\def\BibTeX{{\rm B\kern-.05em{\sc i\kern-.025em b}\kern-.08em
    T\kern-.1667em\lower.7ex\hbox{E}\kern-.125emX}}
\begin{document}

\title{Efficient Airy Beam Training for Quasi-LoS Terahertz Near-Field Communications }

\author{
    \IEEEauthorblockN{Wenqi Zhao and Chong Han}
    \IEEEauthorblockA{Terahertz Wireless Communications (TWC) Laboratory, 
    Shanghai Jiao Tong University, Shanghai, China. \\
    Email: \{wenqi.zhao, chong.han\}@sjtu.edu.cn}
}
\maketitle

\begin{abstract}
With the enlargement of antenna apertures in 6G Terahertz (THz) communications, the Rayleigh distance expands significantly, rending near-field propagation a dominant scenario in THz links. Beyond conventional Line-of-Sight (LoS) and Non-Line-of-Sight (NLoS) conditions, quasi-LoS scenarios with partial obstructions have emerged as a critical challenge. Airy beams offer a promising solution to circumvent obstacles due to their unique curving trajectory. However, existing Airy beam training methods typically rely on parameter-based sampling or exhaustive search, leading to significant pilot overhead and low training efficiency.
In this paper, an efficient Airy beam training framework is proposed to address this research gap. First, the theoretical bounds of Airy beam generation under finite apertures to prune physically invalid codewords are derived. Based on this, a two-stage Non-Uniform Polar Codebook (NUPC) design is presented, utilizing a probing mechanism to resolve the bending direction and a polar-domain spatial sampling strategy to generate Airy beams. To address ultra-low latency requirements, a Fast-Scanning 1D Codebook (FS1C) is further developed that sweeps the entire LoS region with minimal codewords.
Simulation results demonstrate that NUPC achieves a higher average spectral efficiency (SE) by 13.4~bit/s/Hz while reducing training overhead by 54.2\% compared to the state-of-the-art hierarchical focusing-Airy codebook (HFAC). Furthermore, FS1C reduces overhead by 92.9\% with only a marginal 0.3~bit/s/Hz reduction compared with HFAC.
\end{abstract}


\section{Introduction}

Terahertz (THz) communication has been envisioned as a key technology for sixth-generation (6G) and beyond wireless systems, primarily due to its potential to provide unprecedented multi-GHz bandwidth \cite{Akyildiz-2022-Terahertz}. In the context of hybrid beamforming, the continuous expansion of antenna aperture sizes and the continuous rise in operating frequencies have caused the Rayleigh distance to extend significantly. This shift ensures that transceivers increasingly operate within the near-field region rather than the conventional far-field region \cite{An-2024-Near-Field}. Consequently, the wave propagation model must transition from planar to spherical waves, a shift that simultaneously unlocks new degrees of freedom for advanced wavefront shaping. 
In practical deployment, the enlarged array aperture implies that the channel environment is no longer restricted to ideal Line-of-Sight (LoS) or completely obstructed Non-Line-of-Sight (NLoS) conditions. Instead, partially blocked LoS scenarios~\cite{Yuan-2023-Spatial}, quasi-LoS have emerged as a critical new scenarios. This is particularly evident in complex indoor environments, such as wireless data centers, where dense server racks, structural partitions, and antenna arrays substantially increase the probability of partial LoS obstruction~\cite{Zhao-2025-WDC}.
Facing these challenges, conventional LoS beam patterns~\cite{Alkhateeb-2015-Limited,Wu-2023-Multiple} suffer from severe performance degradation under quasi-LoS conditions due to their blockage-unaware design principles. While alternatives such as Intelligent Reflecting Surfaces (IRSs) can mitigate blockage, their deployment typically incurs additional hardware costs, necessitates complex channel state information (CSI) acquisition, and introduces significant phase synchronization overhead.

Recently, the Airy beam, a novel beam pattern originally proposed in the field of optics \cite{Zhan-2020-Propagations}, has garnered widespread attention in the THz band. Airy beams exhibit unique self-acceleration, non-diffraction, and self-healing properties, which maintain high energy concentration over long distances and facilitate a curved trajectory capable of bending around obstacles in quasi-LoS scenarios \cite{Guerboukha-2024-Curving}. In \cite{Chen-2024-Curving}, the authors propose a physics-informed learning-based framework to optimize the phase profile of transmit arrays for Airy beam generation, enabling curved wavefront propagation to circumvent obstacles. In \cite{Zhang-2026-NearFieldAiry}, a near-field Airy beamforming is proposed to enhance obstructed links by jointly exploiting Airy amplitude shaping and near-field phase focusing. Furthermore, a closed-form design for Airy beams is established in \cite{Zhao-2026-Airy}, allowing for the derivation of phase profile parameters based solely on the spatial coordinates of the blockage and the receiver. For Airy beam training, the authors in \cite{Weng-2025-Learning} derive a trajectory approximation and proposed a learning-based method using a DFT codebook to identify optimal configurations under unknown blockage conditions. In terms of practical implementation, the work in \cite{Zhao-2025-WDC} advocates for Airy beams in THz wireless data center and introduces a hierarchical focusing-Airy codebook. Despite these advancements, most existing Airy beam training methods rely on parameter-based sampling or exhaustive searches, which are inherently inefficient and incur large pilot overhead.

Building upon our previous work \cite{Zhao-2026-Airy}, where Airy beam parameters can be directly calculated from geometric coordinates, in this paper we propose a systematic and efficient training framework. First, we derive the theoretical bounds for Airy beam generation under a finite aperture to prune physically invalid codewords and reduce pilot overhead. Second, we introduce a two-stage non-uniform polar codebook (NUPC) design. In the first stage, a probing mechanism is utilized to resolve the bending direction, effectively halving the beam search space. In the subsequent stage, a polar-domain spatial sampling strategy is employed to non-uniformly select candidate waypoint locations and synthesize generate Airy beams. Finally, to further minimize overhead for ultra-low latency requirements, we design a fast-scanning 1D codebook (FS1C) that requires minimal pilot signals. This design is motivated by the non-diffraction and self-acceleration properties of Airy beams, which allow each beam to cover a continuous 1D spatial path. By carefully designing the codebook, the Airy beams can sweep the entire LoS region using a minimal set of codewords, ensuring robust blockage evasion with significantly reduced training latency.

The remainder of this paper is organized as follows. The system model and Airy beam framework are presented
in Sec.~\ref{sec:system_model}. The boundary of Airy beam generation is investigated in Sec.~\ref{sec:Airy_boundary}. The non-uniform polar codebook and fast-scanning 1D codebook are introduced in Sec.~\ref{sec:non-uniform_codebook_design} and Sec.~\ref{sec:fast-scanning 1D codebook}. Simulation results are demonstrated in Sec.~\ref{sec:simulation}. Lastly, we
summarize this paper in Sec.~\ref{sec:conclusion}.
\section{System Model and Analytical Airy Beam Framework}
\label{sec:system_model}
In this section, we consider a narrowband THz communication system operating in near-field quasi-LoS scenarios with uniform linear arrays (ULA). Subsequently, we introduce the Airy beam framework including the trajectory equation and the closed-form design solution, which will serve as the foundation for the Airy beam training design in the following sections.
\subsection{System Model}
We consider a two-dimensional Cartesian coordinate system $(z,x)$, where the signal propagates primarily along the positive $z$-axis. The transmitter (Tx) is equipped with an $N_t$-element ULA deployed along the $x$-axis at $z=0$. The coordinates of the $n^{th}$ Tx element are denoted by $(0, x_t^n)$, where
\begin{equation}
    x_t^n = \left(n - \frac{N_t - 1}{2}\right)d, \quad n = 0, \dots, N_t-1,
\label{eq:tx_coords}
\end{equation}
with $d$ being the antenna spacing, yielding a Tx aperture of $D_t = (N_t-1)d$. Similarly, the receiver (Rx) utilizes an $N_r$-element ULA with an aperture of $D_r = (N_r-1)d$. The Rx array is centered at $(z_r, x_c)$, and the coordinates of its $n^{th}$ element are given by $(z_r, x_r^n)$, where
\begin{equation}
    x_r^n = x_c + \left(n - \frac{N_r - 1}{2}\right)d, \quad n = 0, \dots, N_r-1.
\label{eq:rx_coords}
\end{equation}
Furthermore, a potential blockage is assumed to be located at the spatial coordinate $(z_b, x_b)$ between the Tx and Rx arrays.

\subsection{Airy Beam Framework}

According to Fourier optics~\cite{Nikolaos-2019-Airy}, an Airy beam is generated by imposing a cubic phase modulation on a Gaussian beam. The 1D phase profile is given by~\cite{Chen-2024-Curving,Chen-2025-physics}
\begin{equation}
    \phi_x(x_0) = \frac{1}{3}(2\pi B)^3 x_0^3 - \frac{\pi}{\lambda F} x_0^2 - \frac{2\pi}{\lambda}\sin{\theta} x_0,
    \label{eq:airy_phase}
\end{equation}
where $x_0$ represents the transverse coordinate of the antenna array, and $\lambda$ is the carrier wavelength. The parameter $B$ governs the bending curvature, $F$ denotes the focal length, and $\theta$ represents the steering angle. The cubic coefficient $B$ induces self-acceleration and non-diffraction properties, enabling the main lobe to curve around obstacles. The analytical trajectory $x(z)$ of the Airy beam is derived in \cite{Zhao-2026-Airy} as
\begin{equation}
\label{eq:traj}
     x(z) = -\xi\lambda z B - \sin{\theta}z - \frac{S_R^2-S_I^2}{16\lambda\pi^2B^3}z,
\end{equation}
where $\xi$ is the independent variable of the Airy function, $S_R = \frac{1}{z} - \frac{1}{F}$, $S_I = \frac{\lambda}{\pi\omega_0^2}$, and $\omega_0 = D_t/2$ is the Gaussian beam waist.
Given a target receiver center $(z_r, x_r)$ and a spatial waypoint $(z_b, x_s)$ to bypass the blockage, the optimal parameters $\{B,F,\theta\}$ can be analytically determined via the following closed-form solutions~\cite{Zhao-2026-Airy}:
\begin{equation}\label{eq:closed_form_B}
\begin{split}
    B_{\mathrm{opt}} &= \Bigg\{ -\frac{3 \left( \frac{x_r}{z_r} - \frac{x_s}{z_b} \right)}{16 \lambda \pi^2 \omega_0^2} 
     + \sigma \Bigg[ \left( \frac{3 \left( \frac{x_r}{z_r} - \frac{x_s}{z_b} \right)}{16 \lambda \pi^2 \omega_0^2} \right)^2 \\
    &\quad\qquad + \frac{2}{(2\pi)^6 \omega_0^6} + \frac{3 \left( \frac{1}{z_r} - \frac{1}{z_b} \right)^2}{128 \lambda^2 \pi^4 \omega_0^2} \Bigg]^{\frac{1}{2}} \Bigg\}^{\frac{1}{3}},
\end{split}
\end{equation}
\begin{equation}\label{eq:closed_form_F}
    F_{\mathrm{opt}} = \left[ \frac{1}{2}\left(\frac{1}{z_r}+\frac{1}{z_b}\right) + \frac{ 8\lambda\pi^2 \left(\frac{x_r}{z_r} - \frac{x_s}{z_b}\right) }{ \frac{1}{z_r} - \frac{1}{z_b} } B^3 \right]^{-1},
\end{equation}
\begin{equation}\label{eq:closed_form_theta}
    \theta_{\mathrm{opt}} = \arcsin\Bigg( -\xi_{\mathrm{peak}}\lambda B - \frac{x_s}{z_b} - \frac{ \left( \frac{1}{z_b} - \frac{1}{F_{\mathrm{opt}}} \right)^2 - \left(\frac{\lambda}{\pi\omega_0^2}\right)^2 }{ 16\lambda\pi^2 B^3 } \Bigg).
\end{equation}
Here, $\sigma \in \{+1, -1\}$ determines the bending direction, and $\xi_{\mathrm{peak}} = -1.019$ corresponds to the first maximum of the Airy function, i.e., the center of the main lobe. 
Defining the closed-form parameter mapping as $\mathcal{A}(z_b, x_s; z_r, x_r)$, the transmit phase vector is compactly given by
\begin{equation}
     \mathbf{f} = \mathcal{F}(\mathcal{A}(z_b, x_s; z_r, x_c)),
\end{equation}
where $\mathcal{F}(\cdot)$ maps $\{B, F, \theta\}$ to the array phase profile.


\section{Theoretical Bounds of Airy Beam Generation under Finite Array Apertures}
\label{sec:Airy_boundary}
In this section, we characterizes the theoretical bounds of Airy beam generation under a finite transmit aperture to prune physically invalid codewords and reduce pilot overhead. 

Based on caustic theory, the curved main lobe of an Airy beam is the envelope of a constituent ray family. For a trajectory to be physically realizable, the backward extension of its tangent line at any propagation distance $z$ must intersect the transmit plane ($z=0$) within the array aperture $[-D_t/2, D_t/2]$. Otherwise, aperture truncation severely degrades the caustic structure.
Let $X_{\mathrm{int}}(z) = x(z) - z x'(z)$ denote the tangent intercept on the transmit plane. By substituting the analytical trajectory $x(z)$ and its derivative, the intercept can be explicitly written as
\begin{equation}
    X_{\mathrm{int}}(z) = \frac{\frac{1}{F}-\frac{1}{z}}{8\lambda\pi^2 B^3}.
\label{eq:Xint_closed_form}
\end{equation}
\begin{figure}[t]
    \centering
    \subfigure[Waypoint below the boundary.]{\includegraphics[width=0.7\linewidth]{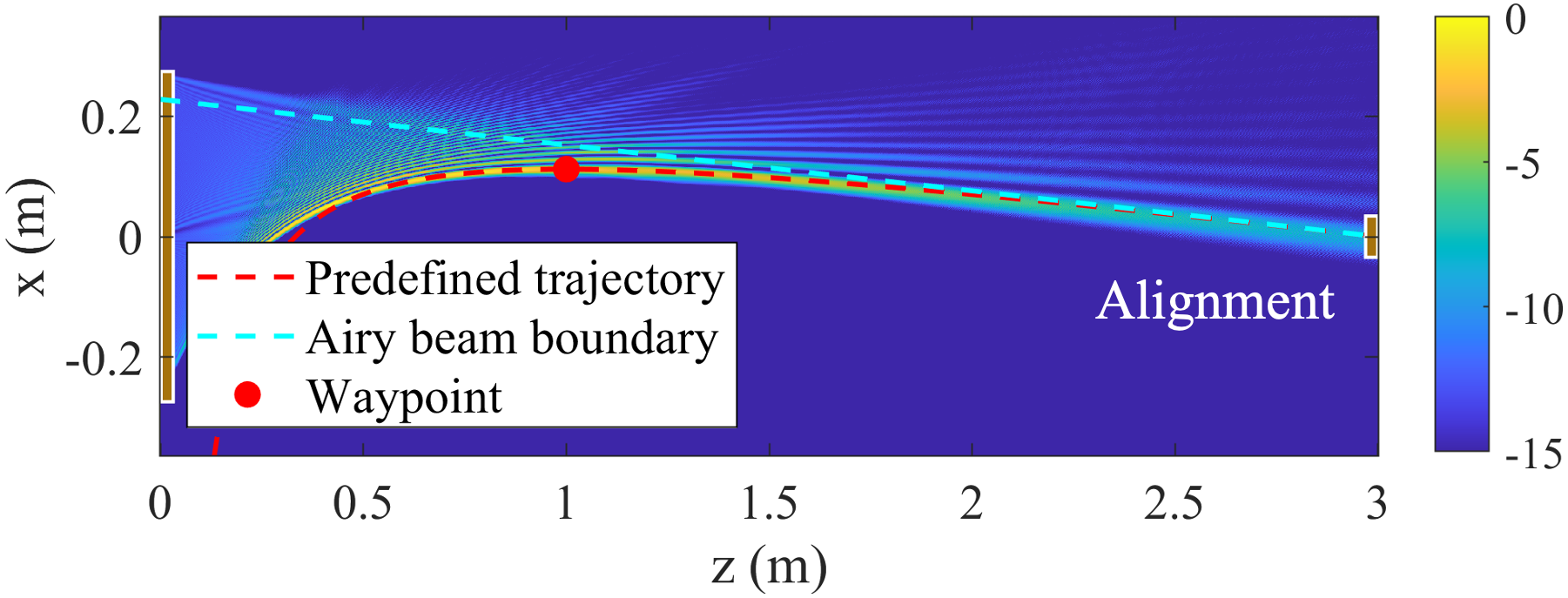}   \label{fig:bb1}}
      \subfigure[Waypoint above the boundary.]{\includegraphics[width=0.7\linewidth]{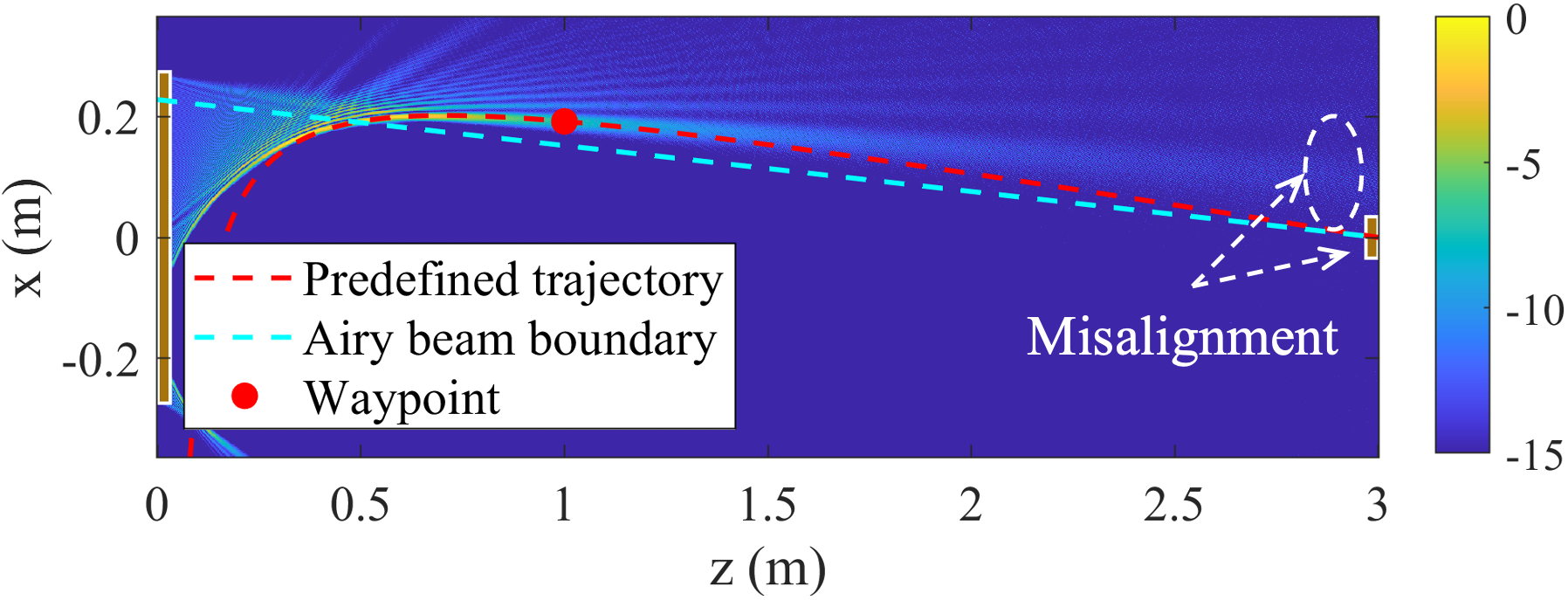}   \label{fig:bb3}}
   \caption{Airy beam boundary ($B>0$).}
   \label{fig:Airy_boundary}
\end{figure}
Since the derivative $\frac{d X_{\mathrm{int}}(z)}{dz}$ is strictly positive, the intercept $X_{\mathrm{int}}(z)$ increases monotonically with $z$. Thus, the most stringent aperture constraint is imposed at the maximum propagation distance, i.e., $z=z_r$. To guarantee physical realizability, the intercept must satisfy $|X_{\mathrm{int}}(z_r)| \leq D_t/2$. By defining the geometric ratios $u = \frac{x_s}{z_b} - \frac{x_r}{z_r}$, $v = \frac{1}{z_b} - \frac{1}{z_r}$, and $X = u/v$, the intercept at the receiver can be rewritten as
\begin{equation}
    X_{\mathrm{int}}(z_r) = X + \frac{v}{16\lambda\pi^2 B^3}.
\label{eq:Xint_zr_uv}
\end{equation}
We next transform the intercept constraint into an explicit geometric boundary in the $(z_b,x_s)$ plane. For a given waypoint and receiver position, substituting the closed-form $B_{\mathrm{opt}}$ into the aperture constraint converts it into a purely geometric condition.
Neglecting infinitesimally small constant terms for analytical tractability, the optimal bending parameter can be approximated as $B_{\mathrm{opt}}^3 \approx v\,M(X)$, where
\begin{equation}
    M(X) = \frac{3X}{16\lambda\pi^2\omega_0^2} + \sqrt{ \left( \frac{3X}{16\lambda\pi^2\omega_0^2} \right)^2 + \frac{3}{128\lambda^2\pi^4\omega_0^2} }.
\label{eq:MX_again}
\end{equation}
Substituting \eqref{eq:MX_again} into \eqref{eq:Xint_zr_uv}, the feasibility condition $|X_{\mathrm{int}}(z_r)| \leq D_t/2$ collapses into a scalar equation solely in $X$. For $B>0$, the upper boundary is attained when the intercept reaches the aperture edge, which can be expressed as 
\begin{equation}
    X+\frac{1}{16\lambda\pi^2 M(X)}=\frac{D_t}{2}.
\label{eq:X_scalar_main}
\end{equation}
By inserting \eqref{eq:MX_again} into \eqref{eq:X_scalar_main} and performing algebraic simplifications, we obtain the critical results $X^\star=\frac{5}{12}D_t.$
Setting $X=X^\star$ and solving for $x_s$ provides the upper feasible boundary for the waypoint, which can be expressed as
\begin{equation}
    x_{s,\max}(z_b) = \frac{5D_t}{12}\left(1-\frac{z_b}{z_r}\right) + x_r\frac{z_b}{z_r}.
\label{eq:xs_boundary_final}
\end{equation}
This equation demonstrates that the feasibility region is bounded by a straight line connecting $(0, \frac{5}{12}D_t)$ and $(z_r, x_r)$. As shown in Fig.~\ref{fig:Airy_boundary}, a waypoint selected below this boundary allows the generated beam to successfully follow the predefined trajectory and reach the receiver. Conversely, a waypoint above the boundary causes aperture truncation and the beam cannot follow the predefined trajectory, resulting in target misalignment.

By symmetry, for the case $B < 0$, the tangent intercept is constrained by the lower edge of the aperture, i.e., $X_{\mathrm{int}}(z_r) \geq -D_t/2$. This yields the symmetric lower feasible boundary:
\begin{equation}
    x_{s,\min}(z_b) = -\frac{5D_t}{12}\left(1-\frac{z_b}{z_r}\right) + x_r\frac{z_b}{z_r}.
\label{eq:xs_boundary_negative}
\end{equation}

\section{Non-uniform polar Codebook Design}
\label{sec:non-uniform_codebook_design}
In this section, we propose a non-uniform polar codebook (NUPC) tailored for Airy beam training. The objective is to efficiently cover the LoS region and bypass blockages while drastically reducing pilot overhead. Our design is divided into two stages: an initial probing mechanism to determine the bending direction, followed by a spatial polar-domain non-uniform sampling strategy that dynamically adapts to the spatial characteristics of Airy beams.

\subsection{Probing Mechanism}

\begin{figure}[t]
    \centering
    \subfigure[Probe beams design.]{\includegraphics[width=0.7\linewidth]{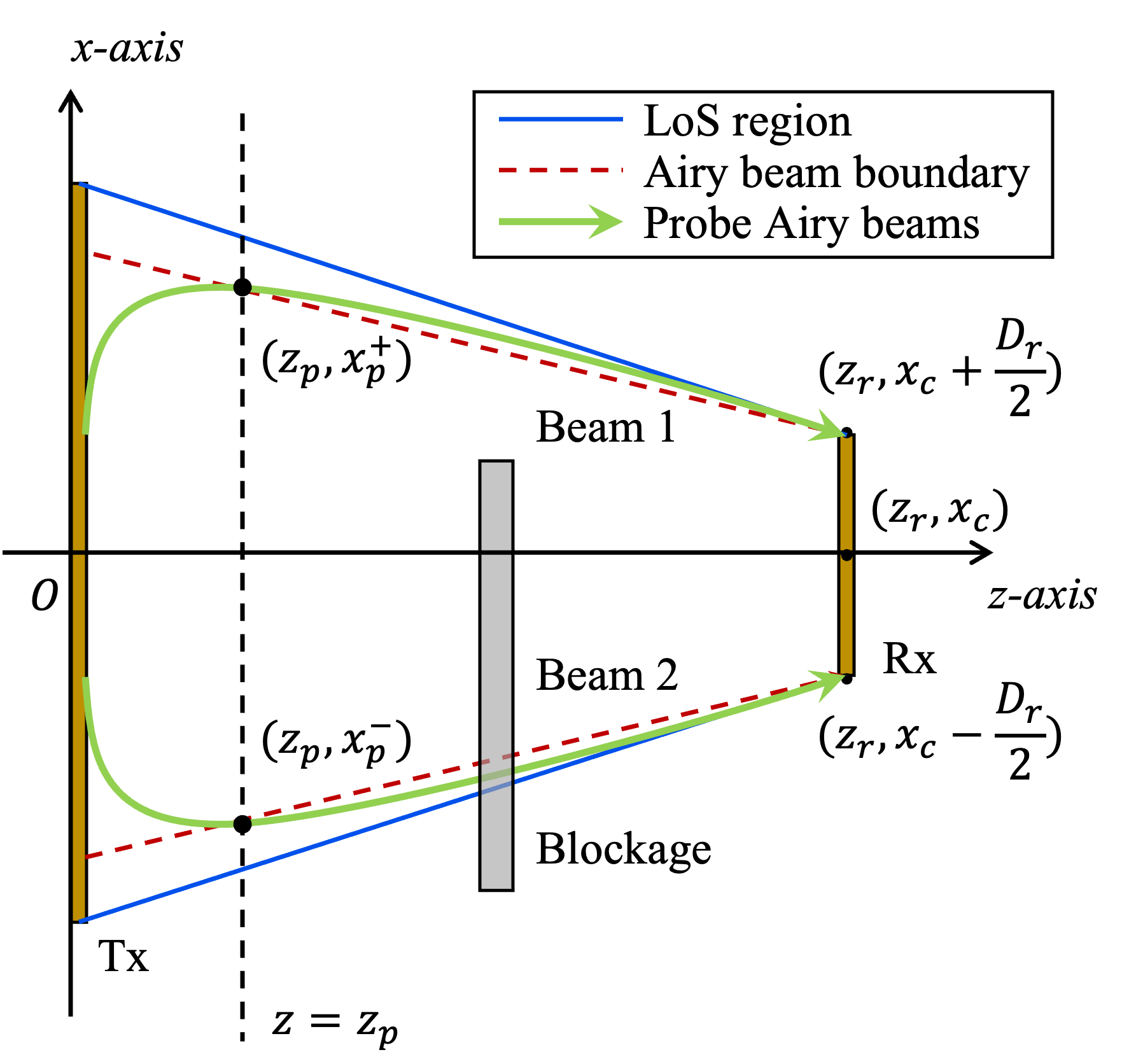} \label{fig:probe_beam} }
      \subfigure[Non-unified polar sampling.]{\includegraphics[width=0.7\linewidth]{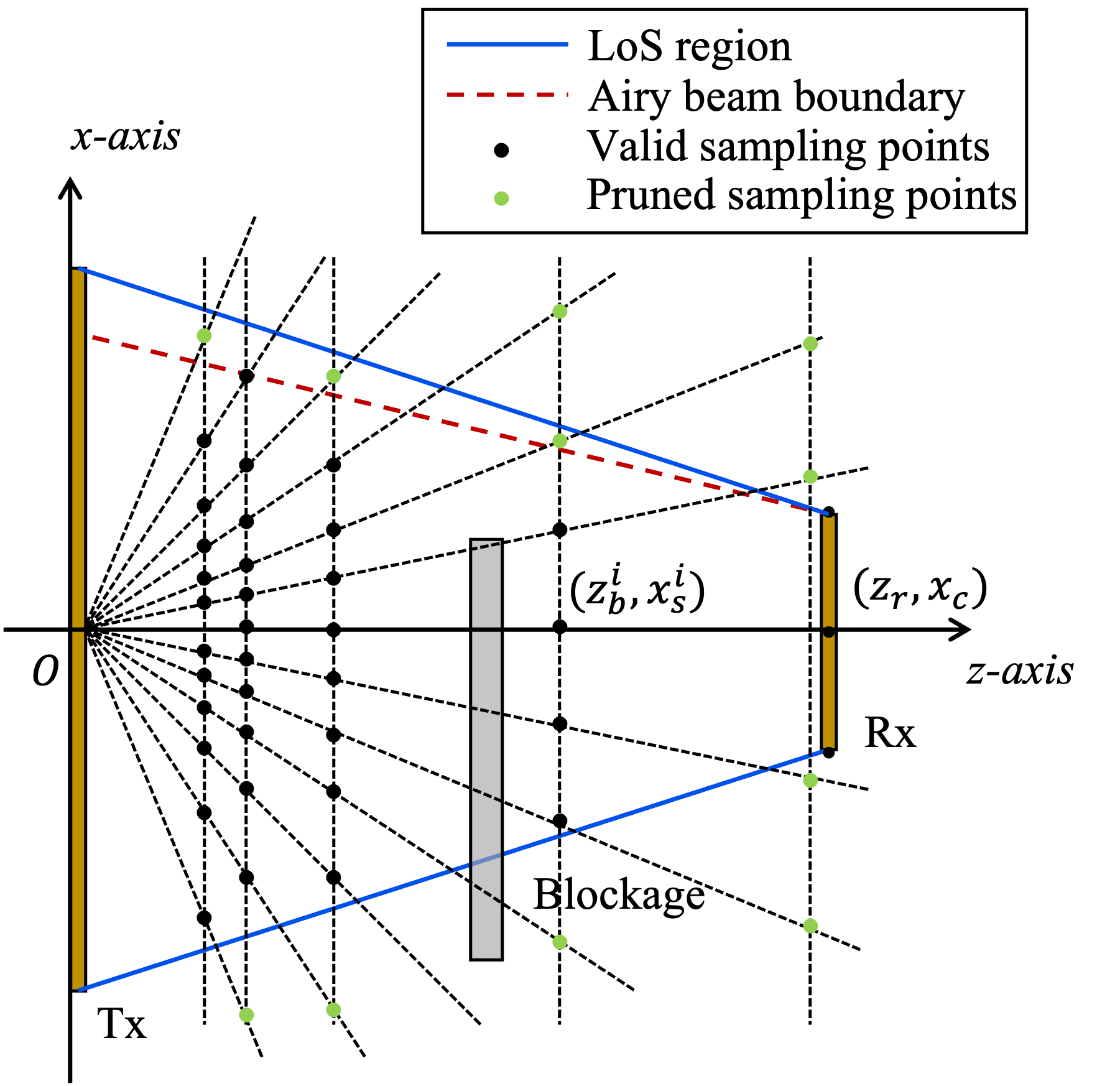}   \label{fig:NUPC}}
   \caption{Non-unified polar codebook.}
\end{figure}

Blindly sweeping the codebook across both upward ($\sigma = 1$) and downward ($\sigma = -1$) bending directions inevitably doubles the training latency. To circumvent this, we introduce a probing mechanism. As shown in Fig.~\ref{fig:probe_beam}, we select a probing plane at $z = z_p$, intersecting the theoretical Airy boundaries derived in Sec.~\ref{sec:Airy_boundary} at upper and lower intercept points, denoted by $(z_p, x_p^+)$ and $(z_p, x_p^-)$, respectively. To maximize spatial coverage, two probe beams are anchored at the physical extremities of the receiver array. Their phase profiles are synthesized as
\begin{equation}
    \mathbf{f}_p^{\pm} = \mathcal{F}\left(\mathcal{A}\left(z_p, x_p^{\pm}; z_r, x_c \pm \frac{D_r}{2}\right)\right).
\end{equation}
The near-field depth $z_p$ is strategically chosen to be close to the transmitter ($z_p \to 0$). This is because the spatial area $S$ enveloped by the probe beam monotonically expands as $z_p$ decreases, reaching its theoretical maximum as $z_p \to 0$. By maximizing $S$, we ensure the greatest possible spatial separation between the probe beams, thereby providing robust blockage discovery in the quasi-LoS path. 
Despite the coverage benefits of a small $z_p$, excessive bending requires a larger $B$, which scatters power into side lobes and reduces main-lobe gain. To navigate this trade-off, $z_p$ is set as the minimum depth ensuring that the received power stays above the detection requirement.

Based on the received energies of the upward and downward probe beams, denoted as $E^+$ and $E^-$ respectively, the optimal bending direction is definitively resolved. In Fig.~\ref{fig:probe_beam}, the downward beam is obstructed while the upward beam survives, yielding $E^+ \gg E^-$. Thus, $\sigma = 1$ is selected, effectively halving the beam search space.

\subsection{Non-uniform Polar Codebook}
Upon fixing $\sigma$, the NUPC is synthesized by systematically sampling potential blockage waypoints. Traditional uniform Cartesian grids are mismatched with the non-linear, curving trajectories of Airy beams. A rigorous inspection of \eqref{eq:closed_form_B} reveals that the required beam curvature is governed by the angular slope $x_s/z_b$ and the inverse distance $1/z_b$. Motivated by this, we propose a coordinate transformation from Cartesian space $(z, x)$ to a polar domain $(\gamma, \phi)$, defined as $\gamma = 1/z$ and $\phi = x/z$.

We perform uniform sampling within the transformed domain $(\gamma, \phi)$. Letting $[z_{\min}, z_r]$ denote the predefined search depth interval for potential blockages, the corresponding inverse distance range is given by $\gamma \in [\gamma_{\min}, \gamma_{\max}]$, where $\gamma_{\max} = 1/z_{\min}, \gamma_{\min} = 1/z_{r}$. Bounded by the LoS region, the angular slope range is $\phi \in [-\phi_{\max}, \phi_{\max}]$, where $\phi_{\max} = \frac{D_r - D_t}{2 z_r} + \frac{D_t}{2 z_{\min}}$. Partitioning this grid into $M \times N$ levels with step sizes $\Delta \gamma$ and $\Delta \phi$, the $i^{th}$ candidate waypoint $(z_b^{(i)}, x_s^{(i)})$ is inversely mapped into the Cartesian domain as:
\begin{equation}
    z_b^{(i)} = \frac{1}{\gamma_{\min} + m \Delta \gamma}, \quad x_s^{(i)} = z_b^{(i)} \left(-\phi_{\max} + n \Delta \phi\right),
\end{equation}
where $m = \lfloor (i-1)/N \rfloor$ and $n = (i-1) \pmod N$.

Taking $\sigma = 1$ as an example, the distribution of the proposed NUPC is illustrated in Fig.~\ref{fig:NUPC}. To minimize overhead, waypoints located outside the LoS region or above the theoretical Airy boundary are pruned. 
For each valid candidate waypoint $(z_b^{(i)}, x_s^{(i)})$, a corresponding receiver point $(z_r,x_r^{(i)})$ must be configured. To ensure the waypoint avoids aperture truncation while maximizing the power, $x_r^{(i)}$ should be strictly bounded. According to \eqref{eq:xs_boundary_final}, the boundary constraint translates to
\begin{equation}
    x_{\text{bound}}^{(i)} = \frac{z_r}{z_b^{(i)}} x_s^{(i)} - \frac{5 D_t}{12} \left( \frac{z_r}{z_b^{(i)}} - 1 \right). 
\end{equation}
Considering the physical boundaries of the Rx array, the optimal target point is selected as
\begin{equation}
    x_r^{(i)} = \max \left( \frac{D_r}{2}, \min \left( x_c, x_{\text{bound}}^{(i)} \right) \right).
\end{equation}
Finally, the comprehensive codebook $\mathcal{C}_{\mathrm{NU}} = \{\mathbf{f}^{(i)}\}_{i=1}^K$ is constructed by substituting these refined parameters into the closed-form design operator:
\begin{equation}
    \mathbf{f}^{(i)} = \mathcal{F} \left( \mathcal{A}(z_b^{(i)}, x_s^{(i)}; z_r, x_r^{(i)}) \right).
\end{equation}
During beam training, the Tx sequentially transmits the codewords in $\mathcal{C}_{\text{NU}}$, and the optimal Airy beam is identified as the one that maximizes the received power at the Rx. 
This NUPC framework ensures that the sampling density dynamically aligns with the physical curvature capabilities of Airy beams, thereby maximizing blockage evasion probability with minimal codeword overhead.

\begin{figure}[t]
    \centering
    \includegraphics[width=0.7\linewidth]{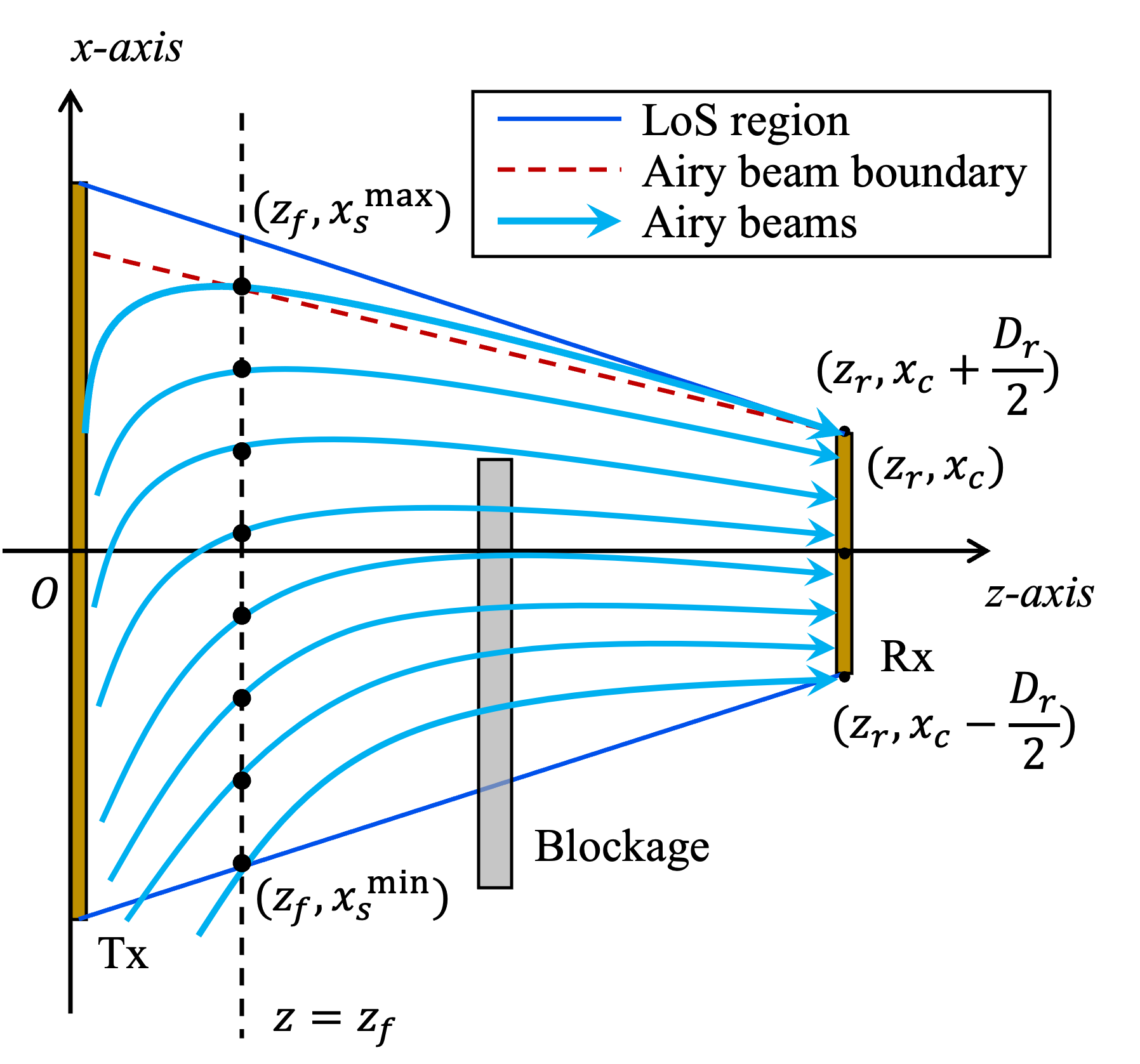} 
    \caption{Fast-scanning 1D codebook.}
\label{fig:FS1C}
\end{figure}

\begin{figure*}[ht]
    \centering
      \subfigure[NUPC.]{\includegraphics[width=0.32\linewidth]{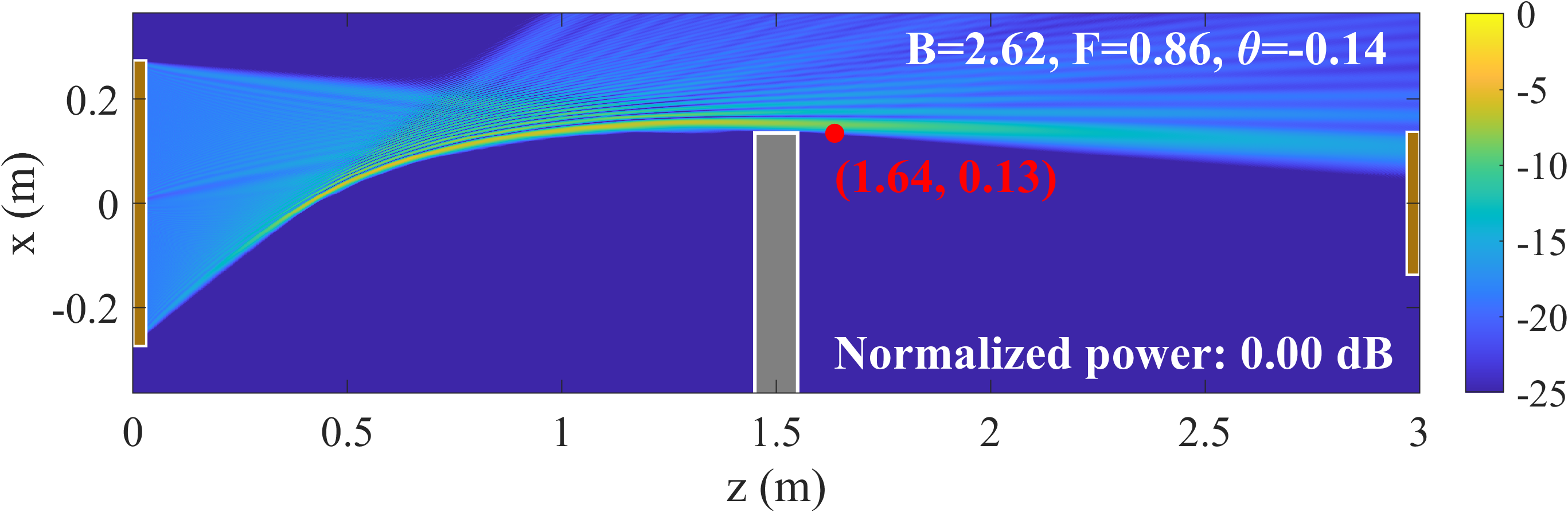}   \label{fig:fig_pol}}
       \subfigure[FS1C.]{\includegraphics[width=0.32\linewidth]{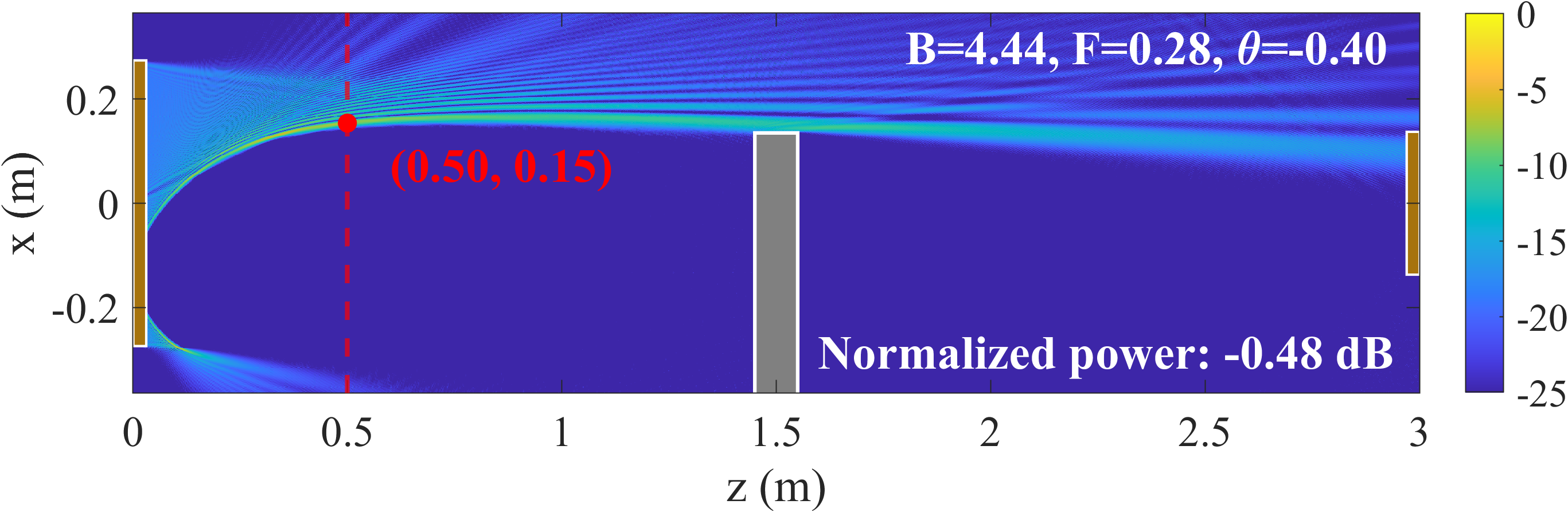}   \label{fig:fig_1D}}
        \subfigure[HFAC\cite{Zhao-2025-WDC}.]{\includegraphics[width=0.32\linewidth]{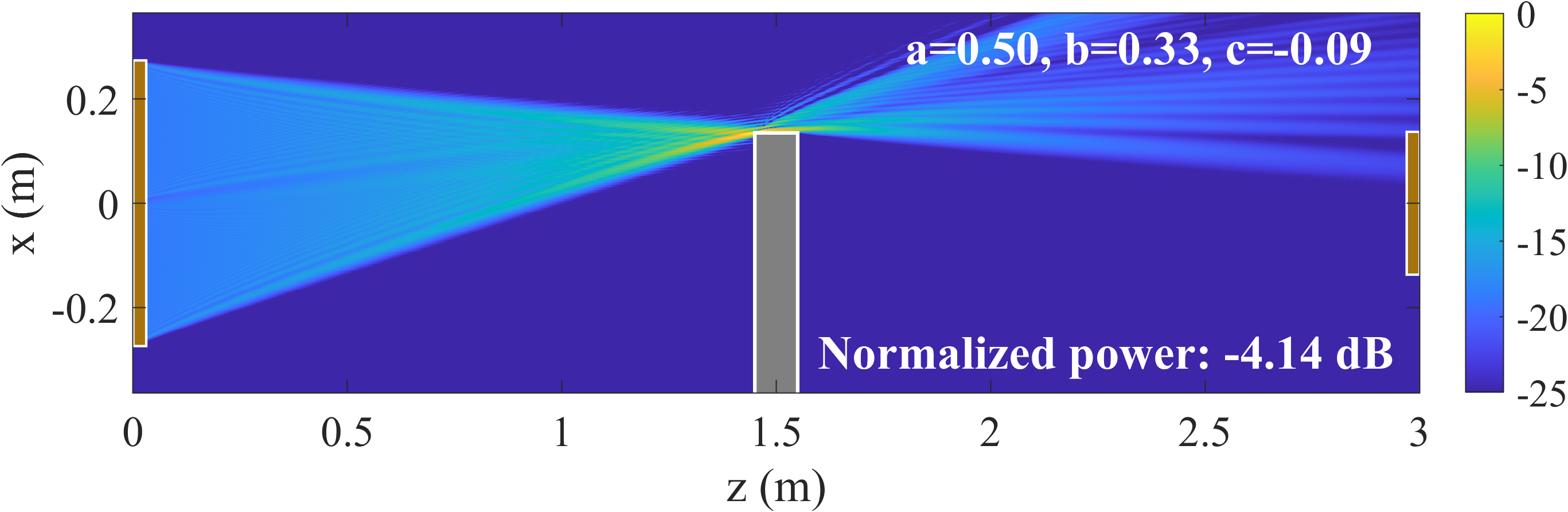}   \label{fig:fig_her}}
   \caption{Performance comparison of various Airy beam codebooks under a quasi-LoS scenario with a 65.8\% blockage ratio.}
   \label{fig:plot_codebook_compare}
\end{figure*}

\section{Fast-Scanning 1D Codebook Design}
\label{sec:fast-scanning 1D codebook}
Although the NUPC significantly compresses the 2D search space, its overhead remains demanding for ultra-low latency applications. To further reduce pilot overhead for ultra-low latency scenarios, we propose a Fast-scanning 1D codebook (FS1C). Unlike conventional beams that concentrate energy at a single focal point, Airy beams distribute energy along a curved trajectory. Because each beam covers a 1D continuous path, a carefully designed sequence of these Airy beams can sweep the entire LoS region using drastically fewer codewords, ensuring high blockage-discovery probability with minimal training latency.

Once the bending direction $\sigma$ is determined via the probing mechanism, we reduce the 2D search space to a 1D parameter sweep. As shown in Fig.~\ref{fig:FS1C}, we fix the waypoint depth near the Tx at $z_f$-plane ensures that the resulting Airy trajectories can reach the extreme boundaries of the LoS region, maximizing the coverage of the codebook.
To reduce the 1D sweep into a deterministic path, we establish a bijective linear mapping between the near-field waypoint $x_s \in [x_{s}^{\min}, x_{s}^{\max}]$ and the target focal point $x_r \in [x_c - \frac{D_r}{2}, x_c + D_r/2]$ on the receiver array, where $x_{s}^{\min}$ and $x_{s}^{\max}$ denote the intersections of $z=z_f$ plane with the LoS lower boundary and the $\sigma=1$ Airy boundary. This dynamic linkage is formulated as
\begin{equation}\label{eq:1D_mapping}
    x_r(x_s) = \left( x_c - \frac{D_r}{2} \right) + \frac{x_s - x_{s}^{\min}}{x_{s}^{\max} - x_{s}^{\min}} D_r.
\end{equation}
By substituting \eqref{eq:1D_mapping} into the closed-form design equations, the entire beam profile is exclusively determined by  $x_s$. 

Next, we demonstrate that this 1D configuration ensures full spatial coverage across the LoS region. Based on our closed-form design, the beam parameters $\{B, F, \theta\}$ are analytical functions of $x_s$ and $x_r$. Since $x_r$ is linearly linked to $x_s$ via \eqref{eq:1D_mapping}, and the trajectory equation in \eqref{eq:traj} consists of elementary algebraic and trigonometric compositions, the mapping $x_s \mapsto x(z; x_s)$ is strictly continuous within the feasible physical domain. Furthermore, the positive linear mapping ensures that an increase in $x_s$ induces a monotonic upward shift of the entire trajectory. By the Intermediate Value Theorem, for any spatial coordinate $X_0$ bounded by the extreme trajectories $x(z; x_{s}^{\min})$ and $x(z; x_{s}^{\max})$, there exists a specific waypoint $x_s^* \in [x_{s}^{\min}, x_{s}^{\max}]$ such that $x(z; x_s^*) = X_0$. This continuity guarantees that the 1D sweep creates an unbroken spatial manifold traversing the entire LoS volume. Consequently, for any potential blockage, there resides at least one Airy trajectory in the 1D codebook that can successfully follow the trajectory to bypass the obstacle and reach the receiver.

Then, we uniformly partition the scanning range $[x_{s}^{\min}, x_{s}^{\max}]$ with a sampling interval $\Delta x_s$. The resulting set of discrete waypoints is defined as
\begin{equation}x_s^{(q)} = x_{s,\min} + q \Delta x_s, \quad q = 0, 1, \dots, Q-1,\end{equation}
where $Q$ denotes the number of 1D codewords. By substituting each $x_s^{(q)}$ and its linked target $x_c(x_s^{(q)})$ into the design operator $\mathcal{A}(\cdot)$, the 1D codebook $\mathcal{C}_{\text{1D}} = \{\mathbf{f}^{q}\}_{i=1}^{Q}$ is synthesized as
\begin{equation}
\mathbf{f}^{(q)} = \mathcal{F}(\mathcal{A}(z_{f}, x_s^{(q)}; z_r, x_c(x_s^{(q)}))).
\end{equation}

During beam training, the Tx sequentially transmits the codewords in $\mathcal{C}_{\text{1D}}$, and the optimal Airy beam is identified as the one that maximizes the received power at the Rx. While the identified beam may not be the strictly optimal solution, it guarantees the fast discovery of a viable propagation path and rapidly provides a robustness communication link, especially in low-latency scenarios.

\section{Simulation Results}
\label{sec:simulation}
In this section, simulations are provided to evaluate the performance of the proposed Airy beam codebooks. The carry frequency is set as 140~GHz. The system employs $N_t=512$ and $N_r=256$ antennas at the Tx and Rx, respectively, with a separation distance of $D=3$ m. The blockage appears in the LoS region resulting a quasi-LoS communication scenario. 

The probing plane for determining the sign of $\sigma$ is positioned at $z_p = 0.5$ m. For the NUPC, the parameters are configured as $\Delta\gamma = 0.221$ m$^{-1}$ and $\Delta\phi = 0.02$. In the case of the FS1C, we set $z_f = 0.5$ m, with a scanning range of $[-0.2509, 0.1901]$ m and a step size of $\Delta x_s = 0.02205$ m. Additionally, for the HFAC \cite{Zhao-2025-WDC}, the sampling intervals are specified as $s_a \approx 0.25$, $s_r \approx 1/3$, and $s_\theta \approx 0.0078$.

Fig.~\ref{fig:plot_codebook_compare} illustrates the beam propagation of various codebooks in a quasi-LoS scenario. Specifically, a blockage with a height of 0.135 m is positioned at $1.5$ m from the Tx, resulting in a 65.8\% blockage ratio. For NUPC, the codebook is constructed by sampling candidate waypoints in the spatial domain to generate corresponding Airy beams, achieving the highest received energy. FS1C performs a rapid 1D spatial search at the reference depth $z_f$, incurring only a marginal $0.48$ dB loss despite the significantly reduced overhead. In contrast, HFAC~\cite{Zhao-2025-WDC} suffers a substantial $4.14$ dB degradation. The fundamental drawback of HFAC is its joint sampling strategy that covers both the spatial coordinates and the cubic phase coefficients. The extra sampling dimension for cubic coefficients not only increases pilot overhead but also limits the identification of optimal beams due to its discretization, decoupling the beam parameters from the actual environment geometry. By focusing the sampling strategy purely on the spatial domain, our proposed codebooks ensure a more direct and effective alignment between the beam trajectory and the blockage location, significantly outperforming the parameter-based search in HFAC.

\begin{figure}[t]
    \centering
    \includegraphics[width=0.7\linewidth]{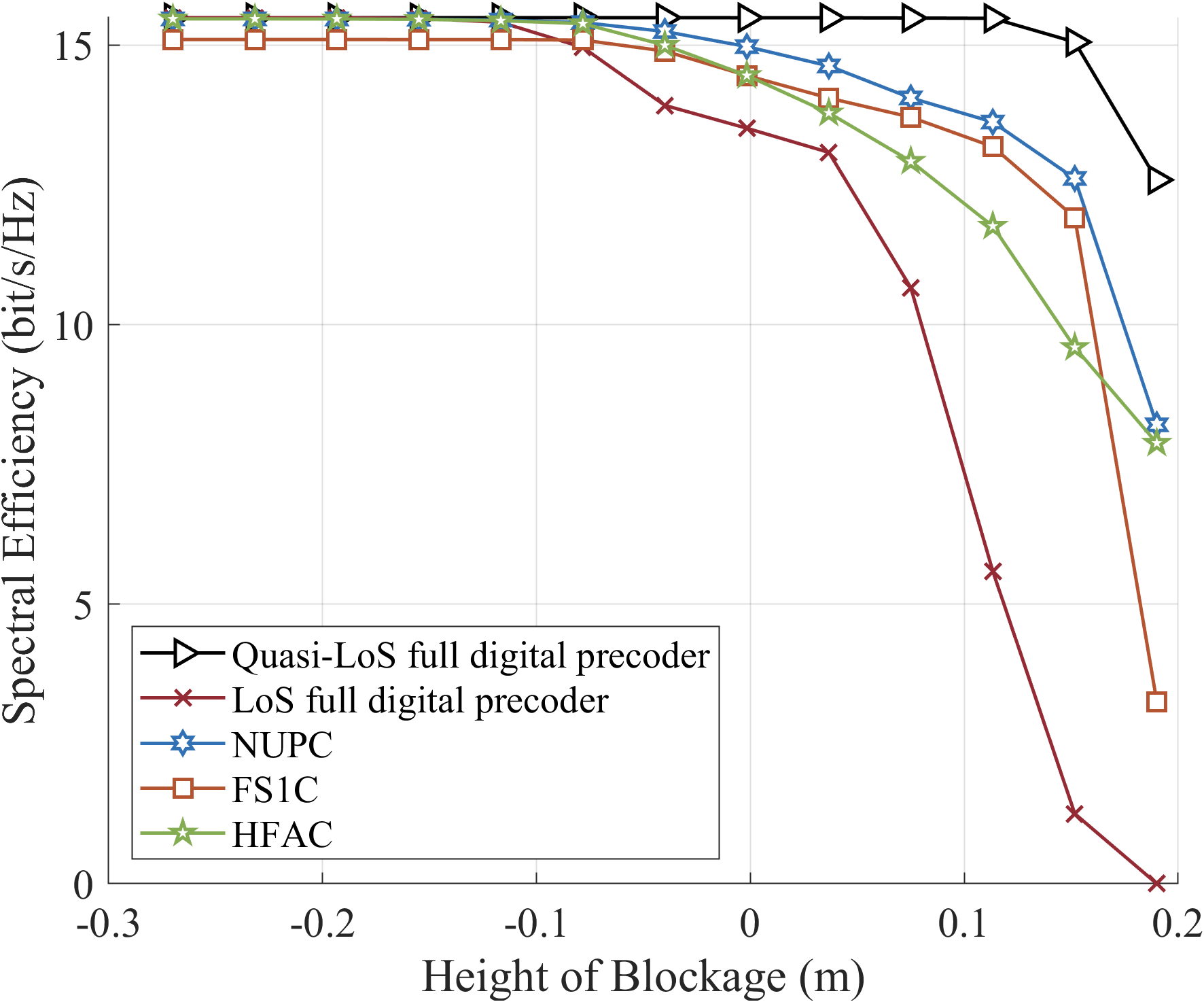} 
    \caption{SE versus height of blockage. $D = 3~\text{m}$, $z_b = 1.5~\text{m}$.}
\label{fig:SE_height}
\end{figure}
Fig.~\ref{fig:SE_height} evaluates the spectral efficiency (SE) against varying blockage heights where the blockage is located at 1.5~\text{m} from the Tx. The quasi-LoS full digital precoder serves as the upper bound at $\approx$ 15.5 bit/s/Hz. In contrast, the conventional LoS-only precoder degrades rapidly, with its SE dropping to nearly 5 bit/s/Hz at $0.113$ m. Our proposed codebooks demonstrate distinct advantages. NUPC maintains high robustness, consistently outperforming HFAC~\cite{Zhao-2025-WDC} across most of the range. FS1C provides a highly efficient alternative. It outperforms HFAC in the low-to-moderate blockage regime and maintains near-optimal SE with only a marginal gap of 0.6~bit/s/Hz compared to NUPC. Although the SE of FS1C falls below HFAC at extreme blockage heights due to the simplified 1D search space, it requires significantly fewer pilots.

\begin{figure}[t]
    \centering
    \includegraphics[width=0.7\linewidth]{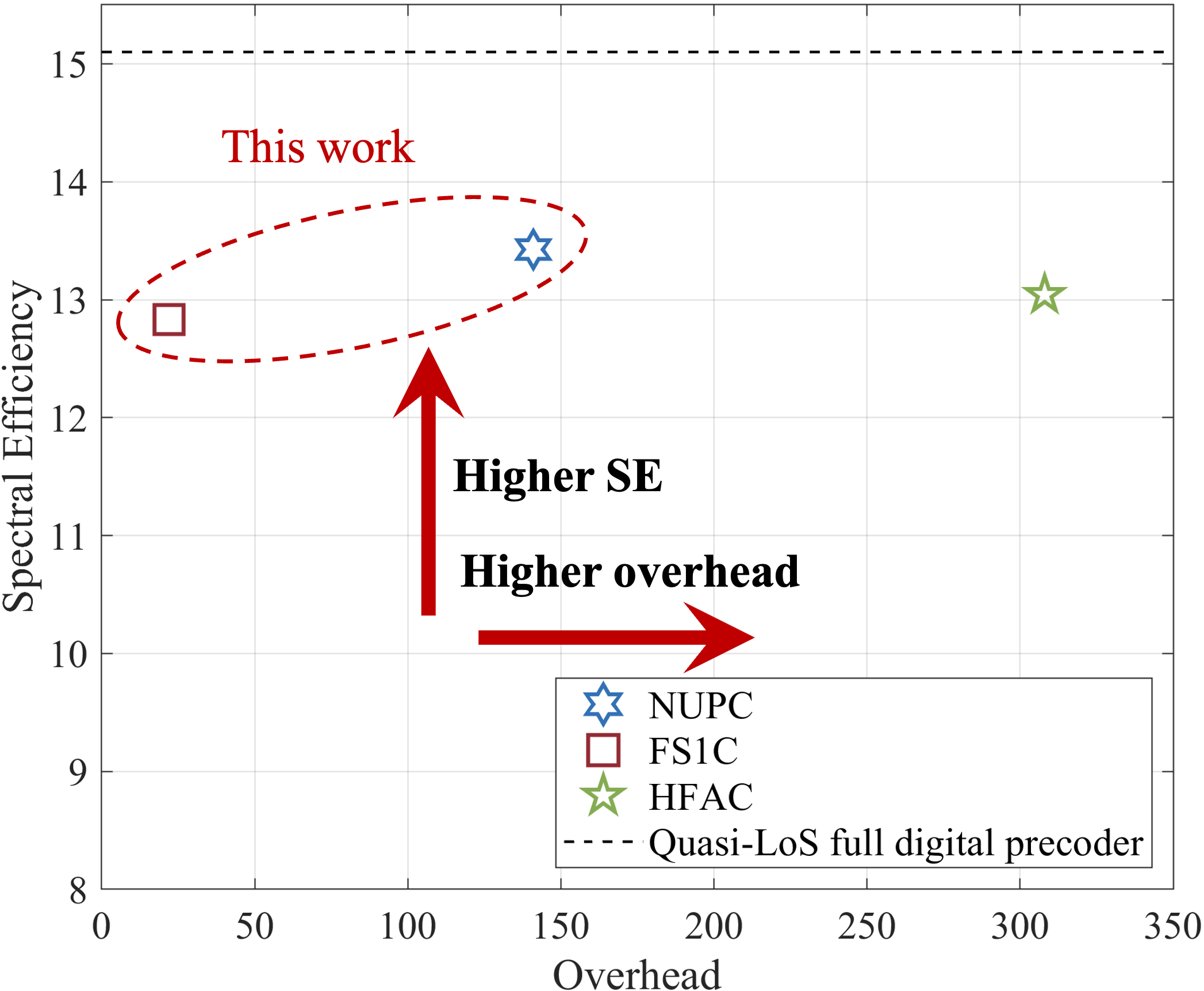} 
    \caption{SE versus overhead.}
\label{fig:SE_overhead}
\end{figure}

To evaluate the overall robustness, Fig.~\ref{fig:SE_overhead} presents the average SE performance across 200 random blockage scenarios within the LoS region. Compared to the HFAC [4], the proposed NUPC achieves a higher average SE of 13.4~bit/s/Hz, while simultaneously reducing the training overhead by approximately 54.2\% compared with HFAC. This gain highlights the efficiency of our spatial-domain sampling, which identifies better curving paths with fewer candidates. Furthermore, the FS1C demonstrates an extreme advantage in low-latency scenarios. It reduces the overhead by a staggering 92.9\% compared with HFAC at the cost of a negligible SE reduction of only 0.3~bit/s/Hz. By shifting from abstract parameter-based searching to geometry-aware spatial mapping, our codebooks achieve fast link recovery with minimal search costs, making them ideal for highly dynamic and obstructed environments.

\section{Conclusion}
\label{sec:conclusion}

In this paper, we have proposed a systematic and efficient Airy beam training framework to address the challenges of quasi-LoS THz communications. By characterizing the theoretical bounds of Airy beam generation under finite apertures, we successfully pruned physically invalid codewords to reduce initial pilot overhead. Then, we introduced a two-stage NUPC that utilizes a probing mechanism to resolve the bending direction and a polar-domain spatial sampling strategy to generate Airy beams. To further satisfy ultra-low latency requirements, we designed the FS1C, which exploits the 1D spatial search of Airy beams to sweep the whole LoS region with minimal codewords. Simulation results demonstrate that NUPC achieves a higher average SE of  13.4~bit/s/Hz while reducing overhead by 54.2\% compared to the state-of-the-art HFAC. Furthermore, FS1C drastically reduces the overhead by 92.9\% with only a marginal 0.3~bit/s/Hz reduction compared with HFAC. These findings reveal that shifting from abstract parameter-based searching to geometry-aware spatial mapping is key to achieving fast link recovery in dynamic and obstructed 6G THz environments.
\bibliographystyle{IEEEtran}

\begin{thebibliography}{10}
\providecommand{\url}[1]{#1}
\csname url@samestyle\endcsname
\providecommand{\newblock}{\relax}
\providecommand{\bibinfo}[2]{#2}
\providecommand{\BIBentrySTDinterwordspacing}{\spaceskip=0pt\relax}
\providecommand{\BIBentryALTinterwordstretchfactor}{4}
\providecommand{\BIBentryALTinterwordspacing}{\spaceskip=\fontdimen2\font plus
\BIBentryALTinterwordstretchfactor\fontdimen3\font minus \fontdimen4\font\relax}
\providecommand{\BIBforeignlanguage}[2]{{%
\expandafter\ifx\csname l@#1\endcsname\relax
\typeout{** WARNING: IEEEtran.bst: No hyphenation pattern has been}%
\typeout{** loaded for the language `#1'. Using the pattern for}%
\typeout{** the default language instead.}%
\else
\language=\csname l@#1\endcsname
\fi
#2}}
\providecommand{\BIBdecl}{\relax}
\BIBdecl

\bibitem{Akyildiz-2022-Terahertz}
I.~F. Akyildiz, C.~Han, Z.~Hu, S.~Nie, and J.~M. Jornet, ``{Terahertz Band Communication: An Old Problem Revisited and Research Directions for the Next Decade},'' \emph{IEEE Trans. Commun.}, vol.~70, no.~6, pp. 4250--4285, Jun. 2022.

\bibitem{An-2024-Near-Field}
J.~An, C.~Yuen, L.~Dai, M.~Di~Renzo, M.~Debbah, and L.~Hanzo, ``{Near-Field Communications: Research Advances, Potential, and Challenges},'' \emph{IEEE Wireless Commun.}, vol.~31, no.~3, pp. 100--107, Jun. 2024.

\bibitem{Yuan-2023-Spatial}
Z.~Yuan, J.~Zhang, Y.~Ji, G.~F. Pedersen, and W.~Fan, ``{Spatial Non-Stationary Near-Field Channel Modeling and Validation for Massive MIMO Systems},'' \emph{IEEE Trans. Antennas Propag.}, vol.~71, no.~1, pp. 921--933, Jan. 2023.

\bibitem{Zhao-2025-WDC}
W.~Zhao, S.~Abadal, G.~Song, J.~Jiang, and C.~Han, ``{Terahertz Wireless Data Center: Gaussian Beam or Airy Beam?}'' \emph{IEEE Trans. Wireless Commun.}, vol.~25, pp. 7922--7938, 2026.

\bibitem{Alkhateeb-2015-Limited}
A.~Alkhateeb, G.~Leus, and R.~W. Heath, ``{Limited Feedback Hybrid Precoding for Multi-User Millimeter Wave Systems},'' \emph{IEEE Trans. Wireless Commun.}, vol.~14, no.~11, pp. 6481--6494, Nov. 2015.

\bibitem{Wu-2023-Multiple}
Z.~Wu and L.~Dai, ``{Multiple Access for Near-Field Communications: SDMA or LDMA?}'' \emph{IEEE J. Sel. Areas Commun.}, vol.~41, no.~6, pp. 1918--1935, Jun. 2023.

\bibitem{Zhan-2020-Propagations}
K.~Zhan, W.~Zhang, R.~Jiao, L.~Dou, and B.~Liu, ``{Propagations of Airy Beams with Quadratic Phase Modulation, and Their Interaction in Paraxial Optical Systems},'' \emph{Opt. Commun.}, vol. 474, p. 126156, Nov. 2020.

\bibitem{Guerboukha-2024-Curving}
H.~Guerboukha, B.~Zhao, Z.~Fang, E.~Knightly, and D.~M. Mittleman, ``{Curving THz Wireless Data Links Around Obstacles},'' \emph{Commun. Eng.}, vol.~3, no.~1, p.~58, 2024.

\bibitem{Chen-2024-Curving}
H.~Chen, A.~Kludze, and Y.~Ghasempour, ``{Curving Around Obstacles via NN-Enabled Wavefront Shaping in Sub-THz Wireless Networks},'' in \emph{Proc. of IEEE Globecom}, Dec. 2024.

\bibitem{Zhang-2026-NearFieldAiry}
S.~Zhang, B.~Di, and L.~Song, ``{Breaking Near-Field Communication Barriers: Focused, Curved, or Airy Beamforming?}'' \emph{arXiv:2604.01704}, Apr. 2026.

\bibitem{Zhao-2026-Airy}
W.~Zhao, C.~Han, and E.~Bj{\"o}rnson, ``{Airy Beam Engineering in Near-field Communications: A Tractable Closed-Form Analysis in the Terahertz Band},'' \emph{arXiv:2603.13866.}, Mar. 2026.

\bibitem{Weng-2025-Learning}
C.~Weng, Y.~Guo, B.~Zhao, Y.~Wang, W.~Chen, and Z.~Li, ``{Learning-Based Blockage-Resilient Beam Training in Near-Field Terahertz Communications},'' \emph{arXiv:2510.25433.}, Oct. 2025.

\bibitem{Nikolaos-2019-Airy}
N.~K. Efremidis, Z.~Chen, M.~Segev, and D.~N. Christodoulides, ``{Airy Beams and Accelerating Waves: An Overview of Recent Advances},'' \emph{Optica}, vol.~6, no.~5, pp. 686--701, May 2019.

\bibitem{Chen-2025-physics}
{H. Chen, A. Kludze and Y. Ghasempour}, ``{A Physics-informed Airy Beam Learning Framework for Blockage Avoidance in Sub-terahertz Wireless Networks.}'' \emph{Nat. Commun.}, vol.~16, no. 7387, Aug. 2025.

\end{thebibliography}

\end{document}